\documentclass[10pt]{ismd08}
\usepackage{graphicx}
\usepackage{cite,./mcite}

\begin{document}
\title{ISMD08 \\ The Ridge, the Glasma and Flow}
\author{Larry McLerran}
\institute{RIKEN-BNL Center and Physics Department, Brookhaven National Lab., Upton, NY, 11973 USA\\  }
\maketitle
\begin{abstract}
I discuss the ridge phenomena observed in heavy ion collisions at RHIC.  I argue that the ridge
may be due to flux tubes formed from the Color Glass Condensate in the early Glasma phase of 
matter produced in such collisions
\end{abstract}

\section{The Color Glass Condensate and the Glasma}

The Color Glass Condensate is the matter which composes the low x part of a hadron wavefunction at high energy.\cite{cgc}  It is  an ensemble of color electric and color magnetic fields which are
in the form of Lienard-Wiechart potentials, $\vec{E} \perp \vec{B} \perp \hat{z}$, where $\hat{z}$ is in the direction of the beam.  

These fields change their character after the collisions, where in a very short time,
$t \sim Q_{sat}^{-1} e^{-\kappa/\alpha_S}$, longitudinal color  electric and magnetic fields are produced.  There are both color electric and magnetic fields because of the duality of QCD under $E \leftrightarrow B$, and the symmetry of the fields in the CGC under this transformation.  Since there is both a longitudinal electric and magnetic field, these fields carry topological charge.  They have a variation in the transverse
size scale of $\Delta r_\perp \sim Q_{sat}^{-1}$, and may be thought of as an ensemble of
flux tubes.  This ensemble of flux tubes is the Glasma.\cite{glasma}

Flux tubes have a long extent in longitudinal coordinate and therefore should have a long range in rapidity.  If a system undergoes 1 dimensional Hubble flow until a time $\tau_F$, and has an extent
in rapidity $y = ln\{(t+z)/(t-z)\}$, then it must have been produced at a time before $\tau_i = \tau_f e^{-y/2}$.  (For example, if $\tau_f$ is a nuclear size, and $y$ is the total rapidity, then $\tau_i$ is the Lorentz contracted nuclear size in the center of mass frame.)

The multiplicity of flux tubes is
\begin{equation}
  {{dN_{FT}} \over {dy}} \sim R^2/R^2_{FT} \sim Q_{sat}^2 R^2
\end{equation}
where $R$ is the radius of the nucleus, and $R_{FT}$ is the radius of the fluxtube.  The flux tube is composed of high intensity color fields, and when these flux tubes decay their multiplicty per flux tube per unit rapidity should be of order $1/\alpha_S$, a factor typical of classical fields.  This yields the familiar Kharzeev-Nardi formula for the multiplicity of initially produced gluons,\cite{kn}
\begin{equation}
{{dN_{gluons}} \over {dy}} \sim {1 \over {\alpha_S}} Q_{sat}^2 R^2
\end{equation}

\section{Flux Tubes and Long Range Rapidity Correlations}

To compute the long range rapidity correlation associated with a flux tube, one needs an expression for the two gluon emission amplitude.  The leading order classical contribution is given by Fig. 1a. \cite{dgmv} Note that if this were emission from purely classical source, then the connected two particle emission amplitude,
\begin{equation}
        {{d^2N} \over {d^2p^1_Tdy^1d^2p^2_Tdy^2}} = \left<{{dN} \over {d^2p^1_Tdy^1}} {{dN} \over {d^2p^2_T dy^2}}\right> -  \left<{{dN} \over {d^2p^1_Tdy^1}} \right> \left< {{dN} \over {d^2p^2_T dy^2}}
        \right>
\end{equation}
\begin{figure}[htbp]
\begin{center}
\begin{tabular} {l l l}
\includegraphics[width=0.15\textwidth] {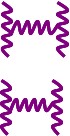}  & &
\includegraphics[width=0.25\textwidth] {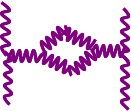} \\
& & \\
a & & b \\
\end{tabular}
\end{center}
\caption{a:  The diagram contributing to the two particle correlation induced by the classical field.
b:  The short range correlation which arises as a quantum correction to the long range classical correlation.}
\label{emission}
\end{figure}
would vanish.
In emission from a Glasma, on the other hand, one has to average the sources, which are treated as a stochastic variable.
If the sources are Gaussian correlated, then one of the contractions of the sources from the nuclei indeed contributes to the disconnected diagram, but there are remaining contributions associated with the different contractions of the sources.\cite{dgmv}  These contributions are associated with quantum interference in the two gluon emission amplitude, and may be difficult to precisely model in simulations which have flux tubes as sources of gluons.  (In experiment, the disconnected contribution is subtracted by the effects of mixed events, since a disconnected contribution corresponds to no event by event correlation.)

In Fig. 1b, the correlated two particle emission amplitude is shown.  This arises as a quantum correction
to the classical emission.  It has a short range correlation in rapidity, and a correction to the long
range term.  (In fact there can in principle be an interference between the classical two particle
emission and the  quantum correction.)
Note that the classical contribution of Fig. 1a, is of order $1/\alpha_S^2$, since it arises from the amplitude squared corresponding to  two classical fields. The quantum correction is of order 1.  

Star has measured the forward backward asymmetry in heavy ion collisions.\cite{srivastava}  The forward backward correlation strength is defined  to be
\begin{equation}
  b = {{<N_FN_B> -<N_F><N_B>} \over {<N_B^2> - <N_B>^2}} 
\end{equation}
If the separation in rapidity for the forward backward correlation is large, the numerator should measure the strength of the long range correlation, but the denominator is a sum over short and long range pieces.  This gives,\cite{mk}-\cite{hidaka}
\begin{equation}
  b = {1 \over {1+ \kappa \alpha_S^2}}
\end{equation}
Since $\alpha_S$ should decrease as centrality increases, this correlation strength should increase as a function of centrality.
\begin{figure}[htbp]
\begin{center}

\includegraphics[width=.99\textwidth] {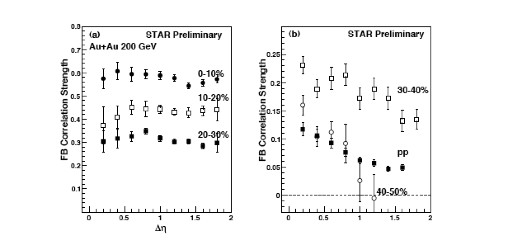} 
\end{center}
\caption{The forward backward correlation strength measured for Gold-Gold collisions in STAR}
\label{starfb}
\end{figure}
The measured strength of the correlation is indeed very strong for central collisions.  It is also claimed that the correlation is much larger than can be expected from impact parameter fluctuations.  

The STAR experiment has also measured directly the two particle correlation.  If one plots the correlation as a function of the azimuthal angle between the two particles, and the rapidity between the two particles, a structure elongated in rapidity and collimated in angle appears.  This is the so called ridge.
It appears either when one or particles have a lower momentum cutoff or when the momenta is integrated over.

In the Glasma description, the two particle correlation should be very strong since essentially all the produced particles arise from flux tubes.  It should also be noted that the two particle correlation predicted by the Glasma extends into the hard regime.  As pointed out by Shuryak, at high momenta, the Glasma flux tubes are the fragmentation jets which appear when one perturbatively computes hard scattering.\cite{shuryak} When two high $p_T$ particles are produced, the charges which induce the scattering are scattered out of the beam, leaving image charges in the fragments of the receding hadrons which produced the jets.  The color fields associated with these image charges are responsible for the beam fragmentation jets, arising from the decay of the flux tube. 

One of the interesting features of the ridge is the collimation in azimuthal angle.  As pointed out by S. Voloshin,\cite{voloshin} this may arise from flow effects.  The position in the transverse plane of the colliding nuclei for a flux tube is localized.  This structure acquires transverse momentum due to radial flow.  The flowing flux tube, when it decays its distribution of decay products will be  preferentially along the direction of motion of the flux tube.  The amplitude of the flux tube distribution will also acquire a non-trivial dependence upon centrality from flow.

With Sean Gavin, we estimated these flow effects for the Glasma using a blast wave model to incorporate the effects of flow.\cite{gavin}  We describe the ridge which is integrated over the energies of both the particles in the two particle correlation function.  Detail of the experimental method for extracting the ridge are given in the Quark Matter 2008 contribution of STAR presented by Daugherity.\cite{daugherity}
\begin{equation}
        {{d^2N} \over {d^2p^1_Tdy^1d^2p^2_Tdy^2}} = \left<{{dN} \over {d^2p^1_Tdy^1}} {{dN} \over {d^2p^2_T dy^2}}\right> -  \left<{{dN} \over {d^2p^1_Tdy^1}} \right> \left< {{dN} \over {d^2p^2_T dy^2}}
        \right>
\end{equation}
\begin{figure}[htbp]
\begin{center}
\begin{tabular} {l l l}
\includegraphics[width=0.50\textwidth] {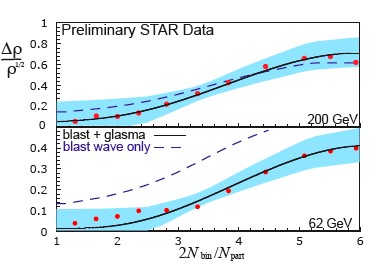}  & &
\includegraphics[width=0.46\textwidth] {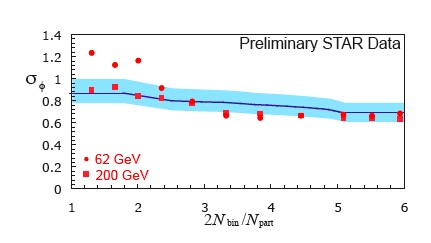} \\
& & \\
a & & b \\
\end{tabular}
\end{center}
\caption{a:  The amplitude for the ridge as a function of centrality at two different energies.  The formula without a factor of $1/\alpha_S$ which arises in the Glasma description is shown as dashed line.  b:  The angular width of the ridge as a function of centrality}
\label{emission}
\end{figure}
In Fig. 3 a, the dependence on the amplitude of the ridge as a function of centrality is shown.  The red points are data.  The solid line is our description.  There is a one parameter ambiguity due to an overall normalization for the emission amplitude, which is not yet computed.    The dashed line curve, is fit to the data not including a factor of $1/\alpha_S$ which arises from the classical field description of the ridge.
In Fig. 3b, the angular width of the ridge is shown as a function of centrality.

At this meeting, Takeshi Kodama informed me that the SPHERIO collaboration from Brazil have seen the ridge in their computations.  This arises because of the string-like configurations in the initial state.  An excellent talk has been presented at the meeting RANP 2008 by Jun Takahashi.\cite{takahashi} 

Clearly the next stage of theoretical analysis of the ridge will be to expand upon the SPHERIO treatment of the ridge.  Some effort need be taken to properly account for the interference diagrams in the two particle correlation function which yield the ridge.  In addition, one needs to understand the effect of energy-momentum conservation, and how it appears in the backward direction from the ridge.

\section{Acknowledgements}  I thank the organizers of the International Symposium on Miltiparticle Dynamics for their kind hospitality.  I also thank Takeshi Kodama for telling me of the recent results from SPHERIO.  I thank my colleagues Adiran Dumitru, Sean Gavin, Francois Gelis, Geroge Moschelli and Raju Venugopalan for their efforts and keen insight during our collaboration.

\end{document}